# Subwavelength resolution for horizontal and vertical polarization by coupled arrays of oblate nanoellipsoids

Carolina Mateo-Segura[1], Constantin Simovski[2], George Goussetis[1] and Sergei Tretyakov[2]

*(1)Heriot-Watt University, Earl Mountbatten Building, Riccarton, London EH14 4AS, United Kingdom*
*(2)Radio Laboratory/SMARAD, Helsinki University of Technology, P.O. Box 3000, FI-02015 TKK, Finland*

A structure comprising a coupled pair of two-dimensional arrays of oblate plasmonic nanoellipsoids in a dielectric host medium is proposed as a superlens in the optical domain for both horizontal and vertical polarizations. By means of simulations it is demonstrated that a structure formed by silver nanoellipsoids is capable of restoring subwavelength features of the object for both polarizations at distances larger than half-wavelength. The bandwidth of subwavelength resolution is in all cases very large (above 13%).

A conventional lens can refocus the far field by means of cancelling the phase that light acquired when propagating away from the source. The resolution in this case is fundamentally restricted for features smaller than half the wavelength. Subwavelength details are contained in evanescent waves, which are characterised by spatial frequencies larger than the free space wavenumber and therefore cannot propagate in the far zone. Based on materials characterised by negative index of refraction[1], a device that can restore subwavelength features in the near field, the so-called superlens, was proposed[2].

However, artificial negative refractive index layers operating as a superlens are very sensitive to losses, spatial dispersion and are extremely narrow-band[3-5]. An alternative technique to produce a superlens for p-polarised light, which only requires negative dielectric permittivity, has also been proposed[2,6]. Metals in the optical range are characterised by negative dielectric permittivity, so that a finely polished (to prevent the scattering of evanescent waves) nanoslab of metal with the permittivity equal but opposite to that of the surrounding media, can be used as a superlens. In this way significant progress was achieved, especially using the tapered multilayer structure called in the modern literature a hyperlens[7].

A different approach to the realization of a superlens is based on the use of a pair of coupled resonant arrays placed in a conventional dielectric matrix, as was theoretically suggested[8] and experimentally demonstrated at microwave frequencies[8,9]. The main advantage of the design is the smaller sensitivity to losses[10]. A proposed implementation of this idea in the optical regime employs metallic nanospheres as the resonating inclusions of two grids supporting surface modes[11]. Enhancement of the near fields, the imaging and subwavelength resolution at the wave distance (more than half-wavelength) from the object was theoretically demonstrated at optical frequencies in coupled arrays of nanospheres[11]. However, the limitation of these devices is that subwavelength imaging can only be achieved when the polarization of the excitation sources is vertical.

In this letter we demonstrate that a pair of arrays consisting of oblate nanoellipsoids can act as a superlens for both polarisations. The structure is composed of two coupled resonant arrays comprising silver particles with oblate ellipsoidal shape, as shown in Fig. 1.

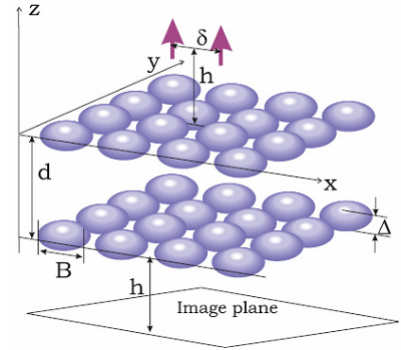

FIG. 1 Schematic representation of the coupled oblate nanoellipsoids arrays structure, proposed to produce subwavelength near-field resolution for both polarisations.

In order to extend the previous analysis[11] to particles of oblate ellipsoidal shape, we replace the scalar polarizability of the sphere by the anisotropic polarizability of the oblate ellipsoid. The break of the geometrical symmetry suggests that the polarizability of an ellipsoid is anisotropic. The inverse polarizability (*j*-th component) of an ellipsoid with permittivity $\varepsilon_m$ and volume $V$ in a host medium with permittivity $\varepsilon_h$ is given by:

$$\frac{1}{\alpha_j} = \frac{\varepsilon_h + (\varepsilon_m - \varepsilon_h)N_j}{V(\varepsilon_m - \varepsilon_h)} - i\left(\frac{k^3}{6\pi}\right) \qquad (1)$$

The geometrical features are captured in the depolarization factor $N_j$. For oblate ellipsoids of eccentricity *e* these are given by (see also Fig. 1):

$$N_z = \frac{1+e^2}{e^3}(e - a\tan(e))$$
$$N_{x,y} = \frac{1}{2}(1 - N_z) \qquad (2)$$

The plasmonic resonance of oblate ellipsoids is polarization dependent and for the field polarized along the *j*-th axis is given by formula:

$$w_r = \frac{w_p}{\sqrt{\left(\frac{1}{N_i}\right)\varepsilon_h + 1}} \qquad (3)$$

Simulations of the electric field distribution were carried out for the structure depicted schematically in Fig. 1 with 23x23 silver ellipsoids in each grid. The host medium is PMMA. The permittivity of silver over the frequency band of our interest has been calculated following[12]. The dimensions of the ellipsoids are $a_x = a_y = 30$ nm and $a_z = 15$ nm, while the grid period is $a = 65$ nm. Such ellipsoids can be with high accuracy modelled as point electric dipoles that gives a strong economy of the computation time compared to the usual discrete dipole approximation (DDA)[13]. The object (source) plane in this example is distanced by $h=a$ from the upper array, while the distance between the two arrays is $d=2a$ and the distance between the object and image planes is $2h+d=4a$. The fields are calculated as described in Ref. 11, both for vertically ($\mathbf{p}=p\mathbf{z}_o$) and horizontally ($\mathbf{p}=p\mathbf{y}_o$) polarised sources. Without any extensive optimization procedure, the suitable frequency of operation is found numerically considering the resolution performance. For this example, this is $0.99f_r$ (where $f_r \sim 5.2650 \times 10^{14}$ Hz) and $1.61f_r$ (where $f_r \sim 3.1809 \times 10^{14}$ Hz) for the vertically and horizontally polarised sources, respectively. Here $f_r$ denotes the resonant frequency of a single ellipsoid for the respective polarization. The effective wavelength in the host medium is equal at these frequencies to $\lambda_{eff} \sim 383.7$ nm and $\lambda_{eff} \sim 390.53$ nm, respectively.

The image is created by the spatial distribution of the co-polarized (with the object polarization) component of the electric field $\mathbf{E}$. We depict the square power of the amplitude of $\mathbf{E}$ normalized to its maximal value in the image plane. The plot in Fig. 2a represents this distribution when the object is a point vertical (z-directed) electric dipole located at ($z=h$, $x=y=0$). Fig. 2b shows the one-dimensional (1D) variant of this distribution – that along the x-axis ($y=0$). The half-power width of the image is equal to $0.164\lambda_{eff}$, which clearly confirms the subwavelength imaging effect at the wave distance from the object (at this frequency the distance between the object and the image plane is equal to $0.63\lambda_{eff}$). Similar plots for the case of a horizontally polarized source are shown in Fig. 3. In this case, the half-power width of the image is $0.251\lambda_{eff}$ and the distance between the object and the image plane equals to $0.672\lambda_{eff}$ ($\lambda_{eff}=390.53$ nm).

To study the resolution of the superlens, the next simulated structure contains two vertical dipole sources located at ($z=h$, $x=a$, $y=0$) and ($z=h$, $x=-a$, $y=0$), corresponding to a subwavelength distance between sources $\delta=0.338\lambda_{eff}$ (see Fig. 1). In Fig. 4a the distribution of the z-component of the electric field in the image plane is depicted and the 1D distribution along x ($y=0$) is shown in Fig 4b. The two point sources are reliably resolved in the image plane. Some additional maxima seen in Fig. 4a (resulting from the eigenmodes of the finite structure) do not degrade the resolution. Similar plots for the case when the object is formed by two horizontal dipoles located at ($z=h$, $x=0$, $y=1.05a$) and ($z=h$, $x=0$, $y=-1.05a$) are shown in Fig. 5. The distance between the source dipoles is also subwavelength (it equals to $0.349\lambda_{eff}$, where $\lambda_{eff}=390.53$ nm). So, the same structure offers imaging and near-field enhancement with subwavelength resolution for objects creating vertically polarized fields (at one frequency) as well as for objects creating horizontally polarized fields (at another frequency).

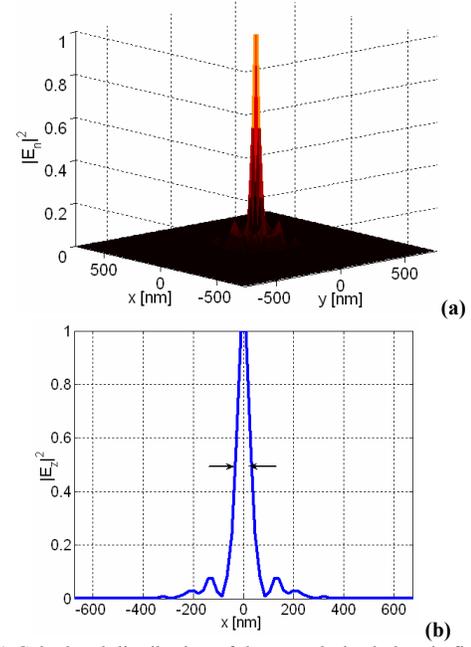

FIG. 2. (a) Calculated distribution of the co-polarized electric field (second power of the amplitude) in the image plane when the object is a single vertically polarized dipole. Fields are normalized to the maximum value. (b) Similar distribution along x ($y=0$) in the image plane.

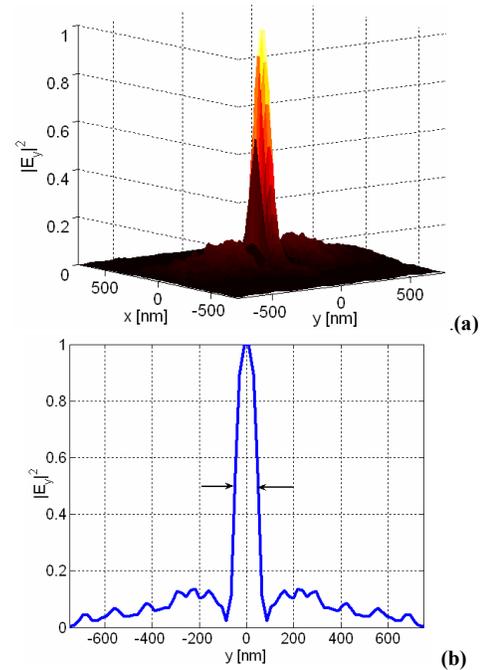

FIG. 3. (a) Calculated distribution of the co-polarized electric field (second power of the amplitude) in the image plane with a single horizontally polarized dipole object. (b) Similar distribution along x ($y=0$) in the image plane.

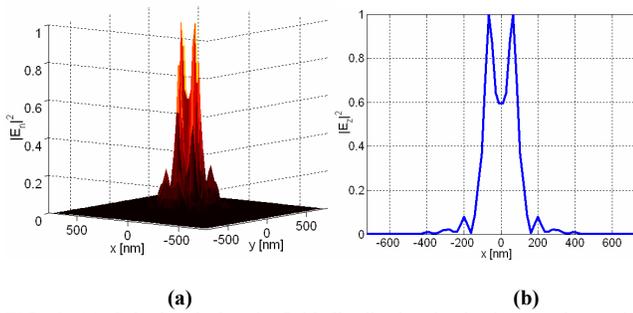

**(a)**         **(b)**

FIG. 4. (a) Calculated electric field distribution in the image plane with two vertically polarized sources. (b) Similar distribution along $y=0$ line in the image plane.

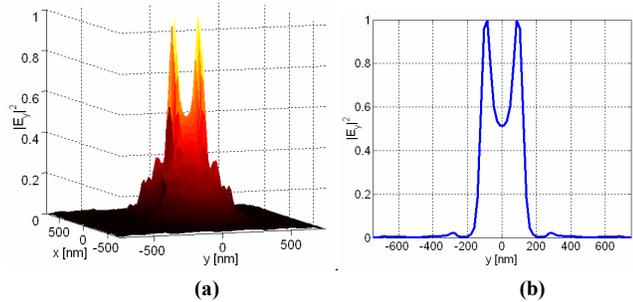

**(a)**         **(b)**

FIG. 5 (a) Calculated electric field distribution in the image plane with two horizontally polarized sources. Fields are normalized to the maximum value. (b) Similar distribution along $x=0$ line in the image plane.

Next, a combination of vertical dipole sources forming an object shaped as "L" (first letter of the word "lens") is considered. The distance between the dipoles was chosen equal to $a$. In order to asses the performance of the superlens, we compare the field in the image plane not with the letter L but with the field in a plane which is distanced only a little ($\lambda_{\text{eff}}/5$) from the object plane calculated in the absence of the superlens. This is more comprehensive comparison since the size of the L-shaped object is comparable with the wavelength and the interference of dipole fields forming the object is visible even in the near zone. The field at the distance $\lambda_{\text{eff}}/5$ from the object plane and the field in the image plane are shown in Figs. 6a and 6b, respectively. The presence of the superlens partially suppresses the interference domain. Though the distance between the object and the image plane is 0.63 $\lambda_{\text{eff}}$, the superlens reliably restores the shape "L" in the image plane.

A thorough investigation of the subwavelength resolution effect has been followed through for different frequencies, not shown here for the sake of brevity. This study revealed that the device produced subwavelength resolution for the vertical polarization between the frequencies $0.982 f_r$ to $0.995 f_r$ and for the horizontal polarization between $1.602 f_r$ to $1.625 f_r$. These correspond to fractional bandwidths of more that 13%.

In summary, in this paper we have demonstrated that a bilayer array of oblate nanoellipsoids is capable of subwavelength resolution in the visible for both polarizations of the object's fields with respect to the image plane – normal polarization and tangential polarization. Subwavelength resolution is achieved at different frequencies for each polarization. A typical resolution is $(0.3$-$0.4)\lambda_{\text{eff}}$, however, the resolution better than $\lambda_{\text{eff}}/5$ is achievable for the vertical polarization.

For the given example the frequency band where the subwavelength resolution was obtained at the wave distances has the relative width more than 13%. The demonstrated advantages of the superlens from oblate ellipsoids compared to spheres studied in[11] become more important in view of nanofabrication issues. Oblate ellipsoids of the present paper correspond to silver nanotablets from which bilayer arrays with required dense package of particles can be prepared using the electronic beam[14] or nanoimprint[15] lithographic methods.

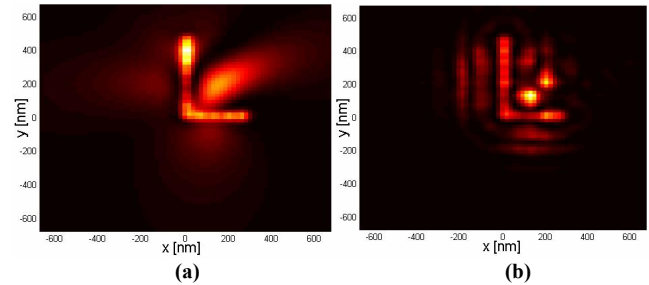

**(a)**         **(b)**

Fig 6 (a) Bird side view of the electric field distribution very close to the object plane (distance=65 nm), when dipoles forming the object are vertically polarized and the distance between these dipoles is 65 nm. (b) Bird side view of the electric field distribution in the image plane (distance from the object plane is equal to $0.63\lambda\text{eff}$).